\documentclass[fleqn,10pt]{wlscirep}
\usepackage[utf8]{inputenc}
\usepackage[T1]{fontenc}
\usepackage{hyperref}
\hypersetup{
  colorlinks = true,
  citecolor = blue,
  urlcolor = blue
}

\usepackage{braket}
\usepackage{graphicx}
\usepackage[mathscr]{euscript}
\usepackage{amsmath}
\usepackage{amsfonts}
\usepackage{bm}
\usepackage{array}
\usepackage{float}

\newcolumntype{x}[1]{>{\hfil$\displaystyle} p{#1} <{$\hfil}}
\newcommand{{\bfL}}{\mbox{\boldmath$L$\unboldmath}}
\newcommand{{\bfl}}{\mbox{\boldmath$l$\unboldmath}}
\newcommand{{\bftheta}}{\mbox{\boldmath$\theta$\unboldmath}}
\newcommand{{\bfdelta}}{\mbox{\boldmath$\delta$\unboldmath}}
\newcommand{{\bfeta}}{\mbox{\boldmath$\eta$\unboldmath}}
\newcommand{{\bfbeta}}{\mbox{\boldmath$\beta$\unboldmath}}
\newcommand{{\bfphi}}{\mbox{\boldmath$\phi$\unboldmath}}
\newcommand{{\bfrho}}{\mbox{\boldmath$\rho$\unboldmath}}
\newcommand{{\bfcalB}}{\mbox{\boldmath$\cal B$\unboldmath}}
\newcommand{{\bfcalE}}{\mbox{\boldmath$\cal E$\unboldmath}}
\newcommand{{\bfcalJ}}{\mbox{\boldmath$\cal J$\unboldmath}}
\newcommand{{\bfcalO}}{\mbox{\boldmath$\cal O$\unboldmath}}
\newcommand{{\bfA}}{\mbox{\boldmath$\emph{A}$\unboldmath}}
\newcommand{{\bfB}}{\mbox{\boldmath$\emph{B}$\unboldmath}}

\def\v#1{{\bf#1}}

\DeclareMathAlphabet\mathbfcal{OMS}{cmsy}{b}{n}

\title{\LARGE{A note on the relativistic temperature}}
\author[*]{\normalsize{Jos\'e A. Heras}}
\affil[*]{Universidad Nacional Aut\'onoma de M\'exico, Ciudad de M\'exico 04510, M\'exico }

\author[**]{\normalsize{ Mar\'ia G. Osorno}}
\affil[**]{Departamento de Ingenier\'ia Qu\'imica y Bioqu\'imica, Instituto Tecnol\'ogico de Toluca, Av. Tecnol\'ogico S/N, Col. Agr\'icola Bellavista, C.P. 52149, Metepec, M\'exico
 }

\begin{abstract}
We show that the largely debated Planck-Einstein and Ott-Arzelies relativistic transformations of temperature do not satisfy the closure group property that two successive temperature transformations must be equivalent to a single temperature transformation of the same form with the involved reference frame velocities satisfying the velocity addition law of special relativity. We then suggest relativistic transformations of temperature that do satisfy this closure requirement and argue that they may be interpreted as particular cases of the so-called directional temperature.
\end{abstract}
\begin{document}

\flushbottom
\maketitle

\thispagestyle{empty}
\section{Introduction}
{\large The idea of a relativistic transformation of temperature has been the subject of a long and heated debate, which has been documented in several
review papers \cite{1,2,3,4,5,6,7,8,9,10}. Different points of view about what is the temperature of a body moving with a constant velocity have been proposed over the years. The first point of view was suggested by Planck \cite{11,12} and Einstein \cite{13}. According to these authors,
the relativistic transformation of the temperature $T$ of a body measured in the rest frame and the temperature of this body $T'$ observed in a frame where the body has the velocity $v$ is given by $T'=T/\gamma$ where $\gamma=1/\sqrt{1-(v/c)^2}$ and $c$ is the speed of light. This formula states that the moving body would appear colder. The Planck-Einstein formula has been supported by many authors \cite{14,15,16,17,18,19,20,21,22,23}. A second point of view was suggested by Ott \cite{24} and Arzelies \cite{25}. According to these authors, the relativistic transformation of temperature is given by the formula $T'=\gamma T$, which states that the moving body would appear hotter. The Ott-Arzelies formula has also been supported by many  papers \cite{25,26,27,28,29,30,31,32}. A third point of view was suggested by Landsberg in a series of papers \cite{33,34,35,36} starting in 1966. He
proposed that the temperature is a Lorentz invariant quantity $T'=T$. This point of view has been supported by some authors \cite{37,38,39,40,41,42}.
It is easy to see that the Planck-Einstein, Ott-Arzelies and Landsberg points of view may be expressed by the generic formula $T'=\gamma^\alpha T$, according to which if $\alpha=-1$ then $T'\!=T/\gamma$, if $\alpha=1$ then $T'\!=\gamma T$ and if $\alpha=0$ then $T'\!=T$.

According to the fourth point of view, the idea of a relativistic transformation of temperature is meaningless because one cannot define a unique temperature in a moving frame, or equivalently expressed, the concept of temperature is well-defined only in a rest frame and therefore there is no Lorentz transformation for temperature. Suggested by Marshall \cite{43} in 1965 this fourth point of view has been supported by many authors \cite{44,45,46,47,48,49,50,51,52,53,54,55,56,57,58,59,60,61,62,63}. In the fifth point of view, the relativistic transformation of temperature is given by $T'=T/[\gamma(1-\beta\cos\theta)],$ where $\theta$ is in the direction of the moving frame with respect to the direction of motion in the rest frame. This relativistic transformation of temperature, which really constitutes a family of transformations depending on the angle $\theta$, describes an idea of temperature that has been called effective or anisotropic or directional temperature. This temperature was discussed in the 1960's by various authors \cite{64,65,66,67} in the context of the black body radiation. The Planck-Einstein, Ott-Arzelies and Landsberg relativistic transformations are particular cases of this directional temperature (as we will show in Sec.~7). The sixth point of view deals with covariant formulations of
temperature. In 1939 van Dantzig \cite{68} observed that the quantity $d\theta= KTdt$, where $dt$ is a time differential, is a Lorentz invariant quantity: $d\theta'=KT'dt'=KTdt=d\theta$. In stating this invariance, he considered the relations $T'=T/\gamma$ (Planck-Einstein formula) and $dt'=\gamma dt$ (time dilation). The invariance of $d\theta$ allowed him to define the four-temperature $\vartheta^\mu=u^\mu/(KT)=(c,u^i)/(KT)$, which states that the temperature is essentially given by the inverse value of the time component of the four-temperature. The promising idea of a covariant formulation of thermodynamics based on the concept of four-temperature
was subsequently explored by van Kampen \cite{38}, Israel \cite{55} and Nakamura \cite{69,70}. More recently, Wu \cite{71} has argued in favour of the four-temperature. In any case the abundant literature on an expected relation between relativity and thermodynamics shows that the debate about the relativistic transformation of temperature is far from over.

In this  note we  want to draw attention to a formal aspect that does not seem to have been previously considered in the proposals for relativistic temperature transformations. This aspect deals with the physical equivalence of the inertial frames, which is an alternative way to state the essential content of the principle of relativity. As noted by L\'evy-Leblond \cite{72}, the physical equivalence of inertial frames implies a group structure for the Lorentz transformations. In connection with this point, one would expect that  any consistent relativistic transformation depending on the relative velocity between reference frames, which we will call here the interframe velocity, should be consistent with the equivalence of the inertial frames in special relativity. This requirement is formally considered by showing that the relativistic transformations of a quantity must form a group and in particular that if they depend on the interframe velocity then they must satisfy
the closure property, which states that the successive application  of two relativistic transformations of a quantity must be equivalent to the application of a single relativistic transformation of this quantity and that the velocities of the associated reference frames must satisfy the parallel-velocity addition law of special relativity. We then show that the Planck-Einstein and Ott-Arzelies transformations of temperature satisfy closure properties involving a parallel-velocity addition law different from that of special relativity and therefore these well-known transformations of temperature are inconsistent with the equivalence of the inertial frames in special relativity. Considering two successive generic transformations of temperature and demanding that they are equivalent to a single temperature transformation of the same form with the involved
velocities of the reference frames satisfying the velocity addition law of special relativity, we infer relativistic transformations for temperature, which state that a body should appear colder for an observer moving away from a rest observer (redshifted temperature) and hotter for an observer moving towards a rest observer (blueshifted temperature). Similar transformations  for the redshifted and blueshifted temperatures where suggested by B\'{\i}r\'{o} and V\'{a}n \cite{73} but for a moving body filled of radiation. By assuming the invariance of the general law of ideal gases, we derive the corresponding relativistic transformations for pressure, heat transfer and internal energy. We also show how the derived transformations of temperature and pressure leave invariant the van der Waals equation for a real gas. Finally, we point out that the suggested transformations of temperature  take the same form than those associated to the directional temperature when  $\theta=0$ and  $\theta=\pi$.
\section{Closure property of relativistic transformations}
Let us recall how the closure property is satisfied by the Lorentz spacetime transformations in their standard configuration. Consider three collinear inertial frames of reference $O, O'$ and $O''$ equipped with their corresponding clocks (see Fig.~1).
\begin{figure}[h]
  \centering
  \includegraphics[width=310pt]{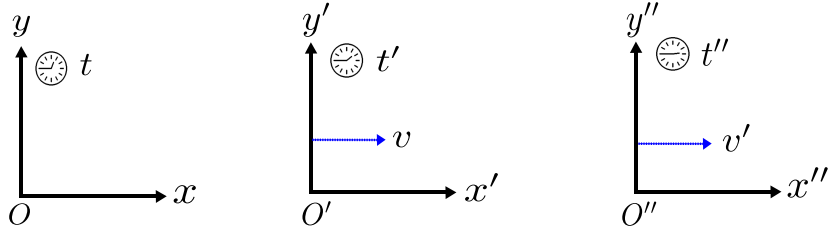}
  \caption{Frame $O'$ moves with velocity $v$ with respect to the rest frame $O$ and frame $O''$ moves with velocity $v'$ with respect to the  frame $O'$. }\label{m}
\end{figure}
The frame $O$ is assumed to be at rest and the frame $O'$ moves with  velocity $v$ with respect to the frame $O$ so that $x'=\gamma(x -v t)$ and $t'=\gamma (t- vx/c^2)$,
where $\gamma=1/\sqrt{1\!-v^2/c^2}$. The frame $O''$ moves with velocity $v'$ with respect to the frame $O'$ so that $x''=\gamma'(x' -v't')$ and $t''=\gamma' (t'- v' x'/c^2)$, where $\gamma'=1/\sqrt{1\!-v'^2/c^2}$. Combining the primed and
double primed transformations we obtain transformations of the same form that  connect the double primed coordinates with the unprimed coordinates:
$x''=\gamma''(x -v''t)$ and $t''=\gamma''(t- v'' x/c)^2$, where
\begin{eqnarray}
\gamma''\!\!\!\!&=&\!\!\!\!\gamma\gamma'\bigg(1+\frac{vv'}{c^2}\bigg),\\
v''\!\!\!\!&=&\!\!\!\!\frac{v+v'}{1+v v'/c^2}.
\end{eqnarray}
This last equation is the well-known parallel-velocity addition law of special relativity. Accordingly, we have verified the closure property that two successive Lorentz transformations in their standard configuration give
another Lorentz transformation (see Fig.~2). Expectably, the relativistic transformations (in their standard configuration) associated to a physical object depending on the interframe velocity should satisfy the closure property of the Lorentz transformations, i.e., they involve the composite quantities $\gamma''$ and $v''$ given by Eqs.~(1) and (2).
\begin{figure}[h]
  \centering
  \includegraphics[width=180pt]{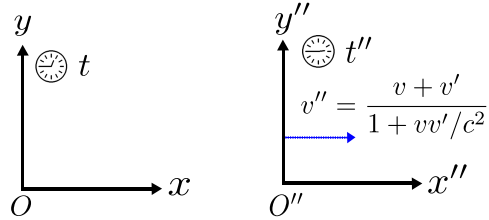}
  \caption{Frame $O''$ moves with velocity $v''=(v+v')/(1+v v'/c^2)$ with respect to the rest frame $O$.}\label{m}
\end{figure}
As an illustrative example, let us consider the case of the electromagnetic field described by its Cartesian components $\v E=(E_x,E_y,E_z)$ and
$\v B=(B_x,B_y,B_z)$. Consider again the three collinear inertial frames of reference $O, O'$ and $O''$ so that the frame $O'$ moves with velocity $v$ with respect to the frame $O$ and the frame $O''$ moves with velocity $v'$ with respect to frame $O'$ as shown in Fig.~1. Henceforth the  unprimed, primed and double primed quantities are respectively measured by the frames $O, O'$ and $O''$. The rest frame $O$ measures the components $E_y$ and $B_z$ and the frame $O'$ measures the components $E_y'$ and $B_z'$.
The unprimed and primed components are connected by the relativistic transformations $E_y'=\gamma(E_y -vB_z/c)$ and $B_z'=\gamma(B_z- vE_y/c)$, where $\gamma=1/\sqrt{1\!-v^2/c^2}$. On the other hand, the frame $O''$ measures the components $E_y''$ and $E_y''$. The primed and double primed components are connected by the transformations $E_y''=\gamma'(E_y' -v'B_z'/c)$ and $B_z''=\gamma'(B_z'- v'E_y'/c)$ where  $\gamma'=1/\sqrt{1\!-v'^2/c^2}$.  Combining the primed and
double primed transformations, we obtain the transformations
$E_y''=\gamma''(E_y -v''B_z/c)$ and $B_z''=\gamma''(B_z- v''E_y/c)$, which connect the components $E_y$ and $B_z$ with the components $E_y''$ and $B_z''$. Here $\gamma''$ and $\beta''$ are given by Eqs.~(1) and (2), which shows that the considered field transformations are consistent with the equivalence of the inertial frames in special relativity. We note that this result is independent from the fact that the components of the involved fields are functions of space and time. In other words, this result is based on a purely algebraic procedure in which the dependencies of the fields with respect to the space and time coordinates do not play any relevant role.

\section{Closure properties of Planck-Einstein and Ott-Arzelies transformations of temperature}

Consider again three collinear inertial frames of reference $O, O'$ and $O''$ in their standard configuration but now equipped with their corresponding thermometers (see Fig.~3).
\begin{figure}[h]
  \centering
  \includegraphics[width=330pt]{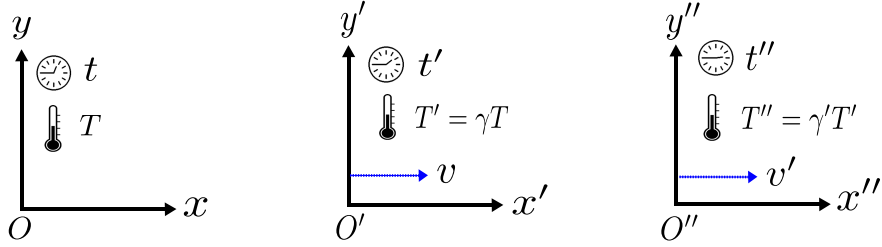}
  \caption{Frame $O'$ moves with velocity $v$ with respect to the frame $O$ and measures the temperature $T'=\gamma T$. Frame $O''$ moves with velocity $v'$ with respect to the  frame $O'$ and measures the temperature $T''=\gamma'T'$. }\label{m}
\end{figure}
The   Ott-Arzelies transformation $T'=\gamma T$ connects the temperature $T$ measured in the rest frame $O$ with the temperature $T'$ measured in the frame $O'$ which moves with velocity $v$ with respect to the frame $O$. Here $\gamma=1/\sqrt{1-(v/c)^2}$. Analogously, the Ott-Arzelies  transformation: $T''=\gamma'T'$ connects the temperature $T'$ measured in the frame $O'$ and the temperature $T''$ measured in the double primed frame $O''$ which moves with velocity $v'$ with respect to the frame $O'$. Here $\gamma'=1/\sqrt{1-(v'/c)^2}$. It follows that $T''=\gamma \gamma' T$ or equivalently $T''=\gamma'' T$ where $\gamma''=\gamma\gamma'$. Now, it can be shown that
\begin{equation}
\gamma''=\gamma\gamma'=\frac{1}{\sqrt{1-v^2/c^2}}\frac{1}{\sqrt{1-v'^2/c^2}}=\frac{1}{\sqrt{1-(v^2+v'^2-v^2v'^2/c^2)/c^2}}
\end{equation}
Therefore the closure property of the  Ott-Arzelies transformation allows us to write the relation $T''=\gamma'' T$ where $\gamma''=1/\sqrt{1-v''^2/c^2}$  with
\begin{equation}
 v''=\sqrt{v^2+v'^2-v^2v'^2/c^2},
 \end{equation}
 which is a velocity addition law different from the law given in Eq.~(2)(see Fig.~4).
 \begin{figure}[h]
   \centering
  \includegraphics[width=230pt]{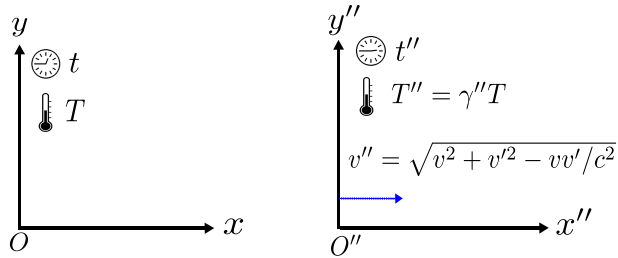}
  \caption{Frame $O''$ moves with velocity $v''\sqrt{v^2+v'^2-v^2v'^2/c^2}$ with respect to the rest frame $O$ and according to the Ott-Arzelies formula measures the temperature $T''=\gamma''T$. }\label{m}
\end{figure}
 This shows that
 the Ott-Arzelies transformation of temperature is not consistent with the equivalence of inertial frames in special relativity which requires
 that $\gamma''$ and $\beta''$ must given by Eqs.~(1) and (2). To further emphasize this result, let us envision the following thought experiment.

Consider again the reference frames $O, O'$ and $O''$ shown in Fig.~1, but now with $v=c/2$ and $v'=c/2$.
Imagine now that a body is attached at the origin of the reference frame $O''$. The velocity addition law in Eq.~(2) states that the body in the frame $O''$ must move with the composite velocity
\begin{equation}
v''= \frac{c/2+c/2}{1+ (c/2)(c/2)/c^2}= 0.80\, c.
\end{equation}
with respect to the frame $O$ (see Fig.~5a).
\begin{figure}[h]
  \centering
  \includegraphics[width=220pt]{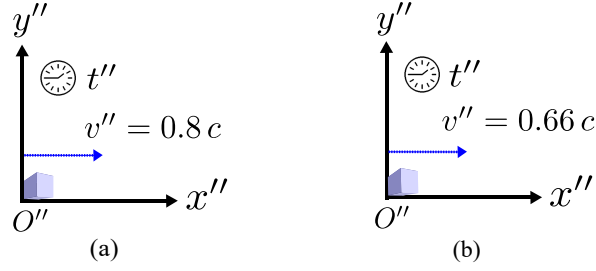}
  \caption{${\bf (a)}$ According to the parallel-velocity addition law of special relativity, the body attached to the frame $O''$
  moves with velocity $v''=0.80\,c$ with respect to the rest frame $O$ and ${\bf (b)}$ according to the parallel-velocity addition law associated to the  Ott-Arzelies formula, the body attached to the frame $O''$ moves with velocity $v''=0.66\,c$ with respect to the frame $O$.}\label{m}
\end{figure}
Nevertheless, according to the composite velocity dictated by the closure property of the Ott-Arzelies transformation of temperature given in Eq.~(4) (i.e. the velocity that is implicit in the factor $\gamma''$ of the transformation $T''=\gamma''T$), the body in the frame $O''$ moves with the composite velocity
\begin{equation}
 v''=\sqrt{(c/2)^2+(c/2)^2- (c/2)^2(c/2)^2/c^2}=0.66\, c,
 \end{equation}
with respect to the frame $O$ (see Fig.~5b). Therefore, the composed velocity $v''=0.8\,c $ of the frame $O''$ predicted by the velocity addition law of special relativity is higher than the velocity $ v''=0.66\,c $ calculated by the velocity addition law associated to  the Ott-Arzelies transformation of temperature! This result is clearly inconsistent or paradoxical to say the least ---both velocities should have the same value because the body is  at rest with respect to the moving observer $O''$.

A similar argument for the Planck-Einstein transformations leads to $T''=T/(\gamma\gamma')$ which can be written as  $T''=T/\gamma''$ where $\gamma''=\gamma\gamma'$. In this case we again have
  $\gamma''=1/\sqrt{1-v''^2/c^2}$ with $v''=\sqrt{v^2+v'^2-v^2v'^2/c^2}$ which is a velocity addition law different from the velocity addition law of special relativity given in Eq.~(2). Thus the Planck-Einstein transformation of temperature is inconsistent with the equivalence of the inertial frames in special relativity.

\section{Redshifted and blueshifted temperatures}
Can one derive relativistic transformations of temperature satisfying the closure property in which the velocities of the reference frames fulfill the rule in Eq.~(2)? In order to address this question, let us consider the hypothetical temperature transformation of the generic form $T'=\gamma {\cal F}(v)T$, where again $T$ is measured in the rest frame $O$ and $T'$ in the frame $O'$ which moves with velocity $v$ with respect to the frame $O$. Here $\gamma=1/\sqrt{1-(v/c)^2}$
and ${\cal F}(v)$ is an unspecified function of the interframe velocity $v$. Similarly, consider the transformation $T''=\gamma'{\cal F}(v')T'$, where $T''$ is measured in the double primed frame $O''$ which moves with velocity $v'$ with respect to the frame $O'$. Here $\gamma'=1/\sqrt{1-(v'/c)^2}$ and ${\cal F}(v')$ indicates that ${\cal F}$ is now a function of $v'$. Combining the primed and double primed temperature transformations we obtain
\begin{equation}
T''=\gamma\,\gamma'{\cal F}(v){\cal F}(v')T.
\end{equation}
Using Eq.~(1) we can write
\begin{equation}
T''=\gamma''\frac{{\cal F}(v){\cal F}(v')}{1+vv'/c^2}T.
\end{equation}
We must now find an explicit form of ${\cal F}$ such that
\begin{equation}
{\cal F}(v'')=\frac{{\cal F}(v){\cal F}(v')}{1+vv'/c^2},
\end{equation}
where $v''$ is given by Eq.~(2). If we were able to find such a ${\cal F}$ then we would have
$T''=\gamma'' {\cal F}(v'')T,$ which would indicate that the successive transformations $T'=\gamma{\cal F}(v)T$ and $T''=\gamma'{\cal F}(v')T'$ produce another transformation of the same form and that the involved velocities  $v,  v'$ and $v''$ satisfy Eq.~(2).

For simplicity, ${\cal F}(v'')$ can be assumed to be a linear function of the form ${\cal F}(v'')=k_1+k_2v''$, where $k_1$ and $k_2$ are constants to be determined. Similarly, ${\cal F}(v')=k_1+k_2v'$ and ${\cal F}(v)=k_1+k_2v$. Thus Eq.~(9) takes the form
\begin{equation}
k_1+k_2\bigg(\frac{v+v'}{1+v v'/c^2}\bigg)=\frac{(k_1+k_2v)(k_1+k_2v')}{1+vv'/c^2},
\end{equation}
where we have used Eq.~(2). Equation (10) implies the values $k_1=1$ and $k_2=\pm 1/c$. Therefore,  ${\cal F}(v)=1\pm v/c, {\cal F}(v')=1\pm v'/c$ and ${\cal F}(v'')=1\pm v''/c$. It follows that the successive transformations $T'=\gamma(1\pm v/c)T$ and $T''=\gamma'(1\pm v'/c)T'$ are equivalent to the transformation $T''=\gamma''(1\pm v''/c)T$ where  $\gamma''$ and $v''$ are given by Eqs.~(1) and (2). From this result we conclude that the relativistic transformations of temperature given by
\begin{equation}
T'=\gamma(1\pm \beta)T,
\end{equation}
where $\beta=v/c$, satisfy the closure group property that two successive temperature transformations are equivalent to a single temperature transformation of the same form with the velocities of the associated reference frames satisfying the velocity addition law of special relativity.

Clearly, the temperature transformations in Eq.~(11) depend not only on the magnitude of the interframe velocity, but also on the sign of this velocity. The idea of temperature transformations depending on the direction of the interframe velocity has previously been explored through the so-called directional temperature which will be briefly discussed in Sec.~7.

Considering the sign of the velocity $v$ in the Lorentz transformations in their standard configuration $x'=\gamma(x -v t)$ and $t'=\gamma (t- vx/c^2)$ it follows that the negative sign of the velocity in Eq.~(11) corresponds to the relativistic temperature transformation in which the frame $O'$ moves along the positive direction of the $x$-axis of the rest frame $O$ (as noted below, this choice of sign is consistent with the formulas of the relativistic longitudinal Doppler effect). Thus,
\begin{equation}
T'=\gamma(1-\beta)T,
\end{equation}
is a relativistic temperature transformation that states that the temperature of a body decreases as measured by an observer that moves away from the rest observer. Analogously,  the positive sign of the velocity in Eq.~(11) corresponds to the relativistic temperature transformation in which the frame $O'$ moves along the negative direction of the $x$-axis of the rest frame $O$. In other words,
\begin{equation}
T'=\gamma(1+\beta)T,
\end{equation}
is a relativistic temperature transformation for temperature that states that the temperature of a body increases as measured by an observer that moves towards the rest observer.

The relativistic transformations in Eqs.~(12) and (13) exhibit the same form  than those of the relativistic Doppler effect: $\nu'_{\small {R}}\!=\gamma(1 -\beta)\nu$ when the source recedes from the rest observer along the line joining observer and source (relativistic redshift) and $\nu'_{\small {B}}\!=\gamma(1 +\beta)\nu$ when the source approaches to the rest observer (relativistic blueshift). We can use the terminology of the Doppler effect and call $T'_{\small {R}}\!=\gamma(1 -\beta)T$ the redshifted temperature and $T'_{{\small B}}\!=\gamma(1 +\beta)T$ the blueshifted temperature. Interestingly, the temperature-frequency relation $T'_{\small R}/\nu'_{\small R}=T/\nu=T'_{\small {B}}/\nu'_{\small B}$
is an invariant quantity. Similar relativistic temperature transformations for the redshifted temperature and the blueshifted temperature were suggested by B\'{\i}r\'{o} and V\'{a}n \cite{73} and also by  B\'{\i}r\'{o} \cite{74}  in their treatment on the relativistic thermodynamics, which is constructed from the point of view of the special relativistic hydrodynamics. These authors have come up with the concepts of redshifted and blueshifted temperatures considering a body filled with radiation while we have arrived at these concepts by considering the formal requirement of the closure group property of a generic temperature transformation consistent with special relativity.

\section{Redshifted thermodynamics: when the observer moves away from the rest observer}

We will first to verify that two successive temperature transformations of the form given by Eq.~(12) are equivalent to a single transformation of the same form whose velocity is given by Eq.~(2). Consider again the collinear inertial reference frames $O,O'$ and $O''$, where the measured temperatures are connected by $T'=\gamma(1 -\beta)T$ and $T''=\gamma'(1 -\beta')T'$, which combine to give the transformation
 \begin{equation}
T''=\gamma\gamma'(1 -\beta)(1 -\beta')T.
\end{equation}
It can be shown that
 \begin{equation}
(1 -\beta)(1 -\beta')= (1+\beta\beta')\bigg(1-\frac{\beta+\beta'}{1+\beta\beta'}\bigg).
\end{equation}
Using this relation Eq.~(14) takes the form
 \begin{equation}
T''=\gamma\,\gamma'(1+\beta\beta')\bigg(1-\frac{\beta+\beta'}{1+\beta\beta'}\bigg)T,
\end{equation}
or more compactly,
  \begin{equation}
T''=\gamma''(1-\beta'')T,
\end{equation}
when Eqs.~(1) and (2) (with $v=\beta c,v'=\beta' c$ and $v''=\beta'' c$) are used. Accordingly, two successive transformations of temperature give
another transformation of temperature of the same form with the involved velocities satisfying Eq.~(2) (see Fig.~6). We then conclude that the transformation  in Eq.~(12) is consistent with the equivalence of the inertial frames in special relativity.
\begin{figure}[h]
   \centering
  \includegraphics[width=250pt]{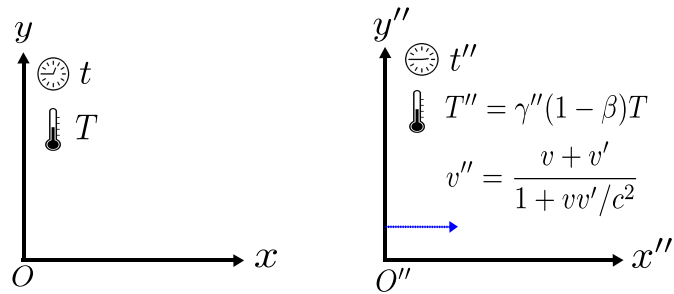}
  \caption{Frame $O''$ moves with velocity $v''=(v+v')/(1+v v'/c^2)$ with respect to the rest frame $O$ and according to the redshifted temperature formula measures the temperature $T''=\gamma''(1-\beta)T$. }\label{m}
\end{figure}

We note that, to a first order of approximation in $\beta$ (i.e., when terms of order $\beta^2$ and greater are ignored and therefore $\gamma \approx 1$), Eq.~(12) becomes
$T'\approx \big(1-\beta\big)T.$
In other words, there is a regime of sufficiently-small velocities, which we will call the slow-velocity regime (defined to be one in which $\beta$ is non-negligible but $\beta^2$ is negligible as compared to 1) in which the transformation   $T'\approx \big(1-\beta\big)T$ holds. This formula states that there is a decreasing of temperature observed in the slow-velocity frame. Notice that $T'\approx \big(1-\beta\big)T$ yields the relation $v\approx |\triangle T|c/T$ where $\triangle T=T'-T$. This relation allows us to obtain the value of the interframe velocity $v$ from the difference of temperatures measured by the slow-velocity observer and rest observer. In the regime of non-relativistic velocities,  in which both $\beta$ and $\beta^2$ are negligible compared to 1, we have $\gamma \big(1-\beta\big)\approx 1$ and then $T'\approx T$ indicating that the temperature does not change for non-relativistic observers. For ultra-relativistic velocities $\beta\to 1$ the temperature measured by the moving observer tends to zero $T'\to 0$. This result can be shown by writing Eq.~(12) as $T'\!=\!\sqrt{(1\! -\!\beta)/(1\!+\!\beta)}\,T$  and observing that limit $\sqrt{(1\! -\!\beta)/(1\!+\!\beta)}\to 0$ as $\beta\to 1$.

We can compare the relativistic transformation of temperature $T'=\gamma(1 -\beta)T$   with the Planck transformation of temperature \cite{12}: $T'_{\small \texttt{Planck}}=T/\gamma$ by considering in this transformation that the moving observer moves away from the rest observer. Both
transformations predict a different decreasing of temperature for the moving observer. By combining both transformations we obtain the relation
$T'=T'_{\small \texttt{Planck}}/(1 +\beta)$, according to which the decreasing predicted by Planck's transformation is lower than that predicted by Eq.~(12).
A comparison of the normalized behaviour of the relations $T'_{\small\texttt{Planck}}/T=1/\gamma$ and $T'/T=\gamma(1 -\beta)$ is shown in Fig.~7.
\begin{figure}[h]
  \centering
  \includegraphics[width=300pt]{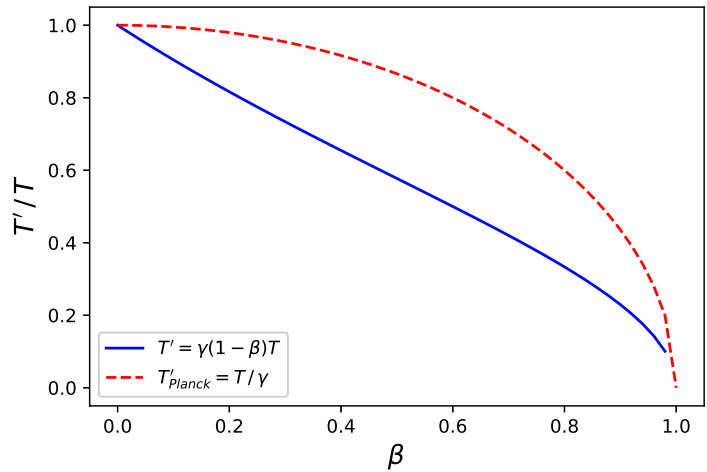}
  \caption{Comparison of the relations $T'/T= \gamma(1-\beta)$ (solid) and  $T'_{\small\texttt{Planck}}/T=1/\gamma$ (dotted).}\label{m}
\end{figure}

To have some sense about the decrease of temperature predicted by $T'=\gamma(1 -\beta)T$, let us consider a hypothetical example at the macroscopic scale. Consider
a stellar object moving with the Hubble's recessional velocity $v = 2\times 10^4$ km/s
and having the temperature $T = 6000K$ in the rest frame $O$.  The Lorentz factor for this hypothetical
object is $\gamma=\sqrt{1-v^2/c^2}\approx 1.002.$ According to the relativistic transformation $T'=\gamma(1 -\beta)T$, the temperature measured in the moving frame $O'$ reads $T' = 5611$ K. The relativistic
correction is about 6.5$\%$

Once we have inferred the relativistic temperature transformation $T'=\gamma(1-\beta)T,$ the next natural step would be to investigate the effect of this transformation in other thermodynamical variables. With this purpose let us consider the general law of ideal gases represented by the equation of state $PV=NKT$, where $P, V$ and $T$ denote the pressure, volume and temperature. The quantities $N$ and $K$ are the number of particles and the Boltzman constant. Both quantities are assumed to be invariant. All unprimed thermodynamical quantities are measured by the rest observer. If we assume the invariance of the law of ideal gases: $P'V'=NKT'$ and the Lorentz transformation for the volume $V'=V/\gamma$, it follows that the  relativistic transformation for pressure associated to the temperature transformation is given by
 \begin{equation}
P'=\gamma^2(1-\beta)P=\frac{P}{1+\beta}.
\end{equation}
This sates that the pressure in a cloud of ideal gas decreases when measured by an observer that moves away from the rest observer.
Using the expansion $(1+\beta)^{-1}\approx 1-\beta +\beta^2 +...$ up to the first order in $\beta$ (the slow-velocity regime), Eq.~(18) can be approximate by $P'\approx \big(1-\beta\big)P,$ which yields $v\approx c|\triangle P|/P,$
where $\triangle P=P'-P$. In the regime of non-relativistic velocities
$1+\beta\approx 1$ and then $P'\approx P$ indicating that the pressure does not change for non-relativistic observers. In the limit of ultra-relativistic velocities ($\beta\to 1$) the pressure measured by an observer that moves away from the rest observer is one half of that measured by the rest observer: $P'=P/2$.

Let us consider the transformations in Eqs.~(12) and (18) in the context of the van der Waals equation for a real gas: $\big(P+aN/V^2\big)\big(V-Nb\big)=NKT$. As is well-known, this equation of state corrects that of ideal gases by considering two properties of real gases: the excluded volume of the particles of gas and the attractive forces between gas molecules.
The parameter $a$ represents the magnitude of intermolecular attraction and the parameter $b$ the magnitude of the excluded volume. The parameters $a$ and $b$ take constant values for a specific gas in the rest frame. But we will see that for the moving frame the quantities $a$ and $b$ will no longer be constant quantities.

When the transformations $T'=\gamma \big(1 -\beta\big)T, V'=V/\gamma$ and $P'=P/(1+\beta)$ are applied to the Van der Waals equation described in the moving frame:
\begin{equation}
\bigg(P'+\frac{a'N}{V'^2}\bigg)\big(V'-Nb'\big)=NKT',
\end{equation}
it preserves its form in the rest frame:
\begin{equation}
\bigg(P+\frac{aN}{V^2}\bigg)\big(V-Nb\big)=NKT,
\end{equation}
whenever the parameters $a$ and $b$ transform as
\begin{equation}
a'=\big(1 -\beta\big)a,\quad b'=\frac{b}{\gamma}.
\end{equation}
The transformation for the quantity $a$ is consistent with the fact that $a'/V'^2$ transforms as a pressure and the transformation for the quantity $b$ describes a relativistic volume transformation.

Considering the transformation in Eq.~(12) we can obtain the corresponding transformation for the heat transfer by assuming the invariance of the relation $dS=\delta Q/T$ among a differential increment of entropy $dS$, an infinitesimal heat transfer $\delta Q$ and a temperature $T$ for reversible processes. If we assume the invariance $dS'=dS$, where $dS'$ is measured by the moving observer and $dS$ by the rest observer and use Eq.~(12) then we get the transformation for the heat transfer:
 \begin{equation}
\delta Q'=\gamma \big(1-\beta\big)\delta Q,
\end{equation}
which states that the heat transfer in a cloud of ideal gas diminishes for an observer that moves away from the rest observer. The relativistic behavior of the heat transfer is the same as the temperature. Thus, in the slow-velocity regime the heat transfer transformation
is given by $\delta Q'\approx \big(1-\beta\big)\delta Q.$
In the non-relativistic limit we have $\delta Q'\approx\delta Q$ and for ultra-relativistic velocities $\delta Q'\to 0$.

From the first law of thermodynamics $dU= \delta Q-PdV,$
where $dU$ is a differential of internal energy, and  $dS=\delta Q/T$, we obtain the relation $TdS=dU+PdV.$
If we again assume $dS'=dS$, Eq.~(12) and $dV'=dV/\gamma$ then we obtain the relativistic transformation for the variation of the internal energy:
 \begin{equation}
dU'=\gamma \big(1-\beta\big)dU,
\end{equation}
which states that the variation of the internal energy of an ideal gas decreases for an observer that moves away from the rest observer. Again, the relativistic behavior of the internal energy is the same as the temperature and the heat transfer. In the slow-velocity regime, this transformation can be approximated by $dU'\approx \big(1-\beta\big)d U.$ In the non-relativistic limit we have $dU'\approx dU$. For ultra-relativistic velocities $dU'\to 0$.

We note that if the definitions of temperature $T=\big(\partial U/\partial S\big)_{V}$   and pressure
$P=-\big(\partial U/\partial V \big)_{S}$
are assumed to be valid in all inertial frames, then we again obtain the relativistic transformations given by Eqs.~(12) and (18),
 \begin{eqnarray}
 T'\!\!\!\!&=&\!\!\!\!\bigg(\frac{\partial U'}{\partial S'}\bigg)_{\!V'}\!\!=\gamma\big(1-\beta\big)\bigg[\bigg(\frac{\partial U}{\partial S}\bigg)_{\!V}\bigg]=\gamma \big(1-\beta\big)T,\\
 P'\!\!\!\!&=&\!\!\!\!-\bigg(\frac{\partial U'}{\partial V'}\bigg)_{\!S'}\!\!=\frac{1}{1+\beta}\bigg[-\bigg(\frac{\partial U}{\partial V}\bigg)_{\!S}\bigg]=\gamma^2(1-\beta)P=\frac{P}{1+\beta}.
\end{eqnarray}
It is pertinent to note that Sutcliffe \cite{26} assumed the Lorentz invariance of entropy, used the definitions $P=-\big(\partial U/\partial V \big)_{S}$ and
$T=\big(\partial U/\partial S\big)_{V}$ and obtained the transformation for temperature proposed by Ott \cite{23}:  $T'=\gamma T$ and a different transformation for pressure: $P'=\gamma^2 P$. These transformations leave invariant the equation of an ideal gas $PV=NKT$ as well as the equation for the variation of entropy $dS= (dU+PdV)/T.$ He interpreted the pressure in the relation  $P'=\gamma^2 P$ as a thermodynamical pressure which was different from the  Lorentz invariant mechanical pressure $P'=P$ appearing in the formulations of Planck \cite{12}, Ott \cite{23} and Landsberg \cite{33}. Similarly, the transformations in Eqs.~(12) and (18) together with $V'=V/\gamma$ leave invariant the equation of an ideal gas and the equation for the variation of entropy. The pressure in Eq.~(18) could also be interpreted as a thermodynamical pressure.

\section{Blueshifted thermodynamics: when the moving observer moves towards the rest observer}
As already noted, the  negative sign in the transformation $T'=\gamma \big(1 -\beta\big)T$ is a consequence of having considered the standard configuration of special relativity in which the frame $O'$ moves along the positive direction of the $x$-axis. But the idea that this frame moves in this direction is a matter of convention not of physical considerations.
One can equally consider that the moving frame moves in the negative direction of the $x$-axis, i.e., that the moving observer moves towards the rest observer. To work in this non-standard configuration  we require to make the replacement $v\to -v$ into the standard configuration (of course, this replacement does not mean that we are obtaining the inverse transformations). After this replacement, we obtain the transformation in Eq.~(13), i.e.,
\begin{equation}
T'=\gamma \big(1 +\beta\big)T,
\end{equation}
which states that the temperature of an ideal gas increases as measured by the observer moving towards the rest observer. Following the same argument used for the temperature transformation $T'=\gamma \big(1 -\beta\big)T$, we can show that the temperature transformation $T'=\gamma \big(1 +\beta\big)T$ is consistent with the equivalence of the inertial frames in special relativity. We can compare Eq.~(26) with Ott's formula \cite{23}: $T'_{\small\texttt{Ott}}=\gamma T$ by considering in this formula that the moving observer moves towards the rest observer. Both transformations predict a different increasing of temperature for an observer moving towards the rest observer. Combining these transformations we obtain the relation
$T'=(1 +\beta)T'_{\small\texttt{Ott}}$, according to which the increasing predicted by Ott's transformation is lower than that predicted by Eq.~(26). A comparison of the relations $T'_{\small \texttt{Ott}}/T=\gamma$ and $T'/T=\gamma(1 +\beta)$ is shown in Fig.~8.
\begin{figure}[h]
  \centering
  \includegraphics[width=300pt]{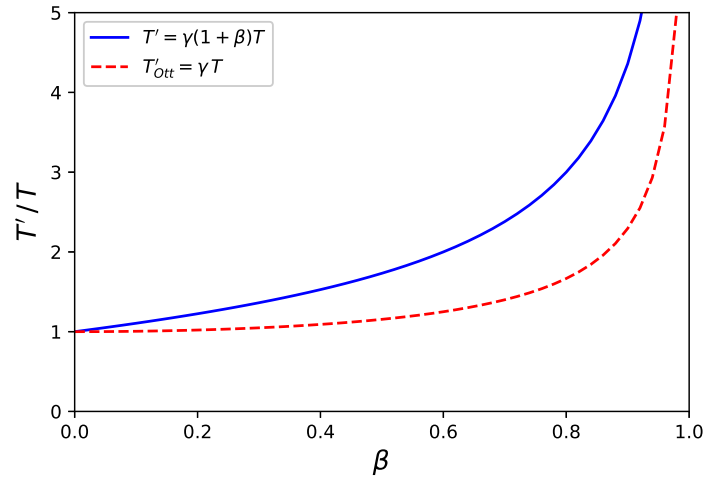}
  \caption{Comparison of the relations $T'/T= \gamma(1+\beta)$ (solid) and  $T'_{\small\texttt{Ott}}/T=\gamma$ (dotted)}\label{m}
\end{figure}

The existence of different relativistic temperature transformations has led some authors to assign
different physical meanings to the temperatures specified in these transformations. For example, Lee and Cleaver \cite{75}
have emphasized the distinction between empirical temperature which arises from the zeroth law of thermodynamics and is a Lorentz invariant quantity $T'=T$ with respect to the absolute temperature which follows from the second law of thermodynamics and, according to these authors, is described by the Ott's relation $T'\!=\gamma T$. The concept of temperature emphasized here and involved in the transformation $T'=\gamma \big(1 +\beta\big)T$, the blueshifted temperature, is formally originated by demanding consistency with respect to the parallel-velocity addition law of special relativity.

Analogously, using Eq.~(26) and the general law of ideal gasses  $PV=NKT$ we obtain the relativistic transformation for pressure
\begin{equation}
P'=\gamma^2(1+\beta)P=\frac{P}{1-\beta}.
\end{equation}
This states that the pressure of the ideal gas increases as measured by the observer moving towards the rest observer. Using the same argument that led to Eq.~(22), we can derive the relativistic transformation for the heat transfer
 \begin{equation}
\delta Q'=\gamma \big(1+\beta\big)\delta Q.
\end{equation}
According to this equation, the heat transfer in an ideal gas increases for an observer that moves towards the rest observer. By the same argument used to obtain  Eq.~(23), we can obtain the  transformation for the variation of the internal energy
 \begin{equation}
dU'=\gamma \big(1+\beta\big)dU.
\end{equation}
This states that the variation of the internal energy of an ideal gas increases for an observer that moves towards the rest observer.

\section{Directional temperature}

 Using the identity $\gamma(1 \pm \beta)\gamma(1 \mp \beta)\equiv 1$, Eq.~(11) can be expressed as
\begin{equation}
T'=\frac{T}{\gamma(1\mp\beta)},
\end{equation}
which states that the temperature measured by the moving observer depends on the Lorentz factor $\gamma$ as well as the directional term $\pm \beta$. The relativistic transformations described by Eq.~(30) can be considered as particular cases of the so-called directional, effective or anisotropic temperature defined by the general transformation law given by \cite{64,65,66,67}
\begin{equation}
T'=\frac{T}{\gamma(1-\beta\cos\theta)},
\end{equation}
where $\theta$ is in the direction of the moving frame with respect to the direction of motion in the rest frame.
If $\theta=0$ then Eq.~(31) yields the blueshifted temperature $T'=T/[\gamma(1-\beta)]=\gamma(1+\beta)T$
and if $\theta=\pi$ then it yields the redshifted temperature $T'=T/[\gamma(1+\beta)]= \gamma(1-\beta)T$. However, we should emphasize that while Eq.~(31) arises in the context of the black body radiation, Eq.~(30) has been inferred here via a formal treatment consistent with the equivalence of inertial frames in special relativity. The directional temperature in Eq.~(31) has been discussed in the literature  \cite{64,65,66,67} without having reached a definitive conclusion as to whether or not this angle-dependent temperature has a physical meaning, and more precisely if this temperature can be identified with the classical notion of temperature existing in the standard thermodynamics. In other words, the concept of directional temperature arose from mathematical manipulations connected with the black body radiation and for this reason is unclear whether this temperature has the meaning of a physical temperature or not. For our part, we have shown here that the idea of directional temperature reappears (when  $\theta=0$ and  $\theta=\pi$) by considering the formal requirement of the closure group property of a generic relativistic temperature transformation consistent with the velocity addition law of special relativity. Let us emphasize that Eq.~(31)
represents a family of relativistic transformations for temperature which depend on the parameter $\beta$ and
the angle $\theta$. Some relevant members of this family are:\\

\noindent (a) If $\theta=\pi/2$ then $T'=T/\gamma$,\quad (Planck \cite{12} and Einstein \cite{13})
\vskip 5pt
\noindent (b) If $\theta=-\cos^{-1}(\beta)$ then\quad $T'\!=\gamma T$, \quad (Ott \cite{23}  and Arzelies \cite{24} )

\vskip 5pt
\noindent (c) If $\theta =\cos^{-1}\big(1/\beta-1/(\gamma\beta)\big)$ then\quad $T'\!=T$,\quad (Landsberg \cite{33} )

\vskip 5pt
\noindent (d) If $\theta=0$ then\quad $T'=\gamma(1+\beta)T$,\quad (Eq.~(26))
\vskip 5pt
\noindent (e) If $\theta=\pi$ then \quad  $T'=\gamma(1-\beta)T$.\quad(Eq.~(12))\\

\noindent A comparison  of the normalised behavior of the transformations $(\rm a)-(\rm e)$ is shown in Fig.~9.

\begin{figure}[h]
  \centering
  \includegraphics[width=300pt]{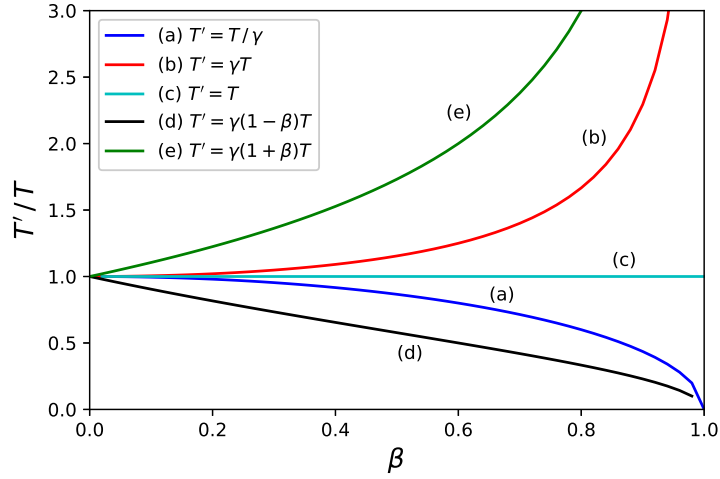}
  \caption{Some relativistic transformations for temperature described by $T'=T/[\gamma(1-\beta\cos\theta)]$. }\label{m}
\end{figure}

\section {Discussion}
Here we have inferred  the relativistic transformations of temperature $T'=\gamma(1\pm\beta)T$ via a formal procedure that did not appeal to any specific thermodynamical relation. Therefore, we were free to choose some specific thermodynamical equations and demand their form invariance to infer the relativistic transformations of those thermodynamical quantities associated to the transformations $T'=\gamma(1\pm\beta)T$. The thermodynamical equations we choose were the general law of ideal gasses $PV=NKT$ and the thermodynamical equation $TdS=dU+PdV$. Accordingly, we demanded the form invariance of these relations:
\begin{equation}
P'V'=NKT', \quad T'dS'=dU'+P'dV',
\end{equation}
and obtained the relativistic transformations given in Eqs.~(18), (22), (23), (27)-(29).

 However, we could have followed a more general procedure. It is clear that the relativistic transformations of thermodynamical quantities like $T, P, S, V, ...$ depend on our initial assumptions. A reasonable program to the formulation of a relativistic thermodynamics can start by considering the well-established relativistic volume transformation $V'=V/\gamma$ and the Lorentz invariance of entropy $dS'=dS$, which is generally accepted, mainly by statistical thermodynamics  considerations (see Gavassino \cite{76} and references therein). We can complete this program by assuming the relations in Eq.~(32). The use of the volume transformation and the invariance of entropy in Eq.~(32) imply the relations $P'V/\gamma=NKT'$ and $T'dS=dU'+P'dV/\gamma,$ whose combination  yields the relation
\begin{equation}
T'\bigg(dS-\frac{NK dV}{V}\bigg)=dU'.
\end{equation}
Four particular cases are relevant,
\vskip 5pt
\noindent (i)$\;\,$ If $T'=T/\gamma$ (Planck-Einstein) then $dU'=dU, P'=P, \delta Q'=\delta Q/\gamma.$
\vskip 5pt
\noindent (ii) $\,\!$ If $T'=\gamma T$ (Ott-Arzelies) then $dU'=\gamma dU, P'=\gamma^2P, \delta Q'=\gamma\delta Q.$
\vskip 5pt
\noindent (iii) If $T'=T$ (Landsberg) then $dU'= dU, P'=\gamma P, \delta Q'=\delta Q.$
\vskip 5pt
\noindent (iv) If $T'=\gamma(1\pm \beta)T,$ (Eq.~(11)) then $dU'= \gamma(1\pm \beta)dU, P'=\gamma^2(1\pm \beta)P, \delta Q'=\gamma(1\pm \beta)\delta Q.$
\vskip 5pt
\noindent As already noted, the temperature transformations in (i) and (ii) are inconsistent with the equivalence of inertial frames in special relativity. The temperature transformation in (iii) does not depend on the interframe velocity and therefore this transformation does not disagree  with the equivalence of inertial frames in special relativity. The temperature transformations in (iv) are consistent with the equivalence of inertial frames in special relativity. In short: special relativity agrees with (iii) and (iv)
but not with (i) and (ii). Equivalently,  we could also have assumed different transformations for $dU'$ and use Eq.~(33) to infer the corresponding transformations for temperature.

However, from purely theoretical considerations nothing does not prevent to consider also the relation $T'dS'=dU'+P'dV'$ together the proposed transformations  $T'=\gamma(1\pm\beta)T$ and the relativistic transformation for the entropy suggested by Avramov \cite{5}: $dS'=dS/\gamma$, would have led us to the relativistic transformations for the pressure $P'=\gamma(1\pm\beta)P$, heat transfer $\delta Q'= (1\pm\beta)\delta Q$ and internal energy $dU'=(1\pm\beta)dU$. Notice that using the transformations $T'=\gamma(1\pm\beta)T$ and  $P'=\gamma(1\pm\beta)P$ and demanding form invariance to the law of ideal gasses $PV=NKT$, we conclude that the Boltzman constant must be transformed as \cite{5}: $K'=K/\gamma$. The idea that the entropy may be not a Lorentz invariant quantity has been discussed by Mares et al \cite{77} and Haddad \cite{78}.

Another interesting case consists in considering the temperature transformations $T'=\gamma(1\pm\beta)T$ and the unusual volume transformations $dV'=\gamma(1\pm\beta)dV$. These two relativistic transformations and the invariance of the thermodynamical relation $TdS=dU+PdV$ allow us  to infer the transformations of entropy $dS'=dS$, pressure $P'=P$, heat transfer $\delta Q'= \gamma(1\pm\beta)\delta Q$ and internal energy $dU'=\gamma(1\pm\beta)dU$. In this hypothetical case, the inferred relativistic transformations leave also invariant the general law of ideal gasses. The interesting point here is that the relativistic transformations  of the involved thermodynamical quantities are consistent with the equivalence of the inertial frames in special relativity. Of course, this case is questionable due to the intervention of the unusual relativistic volume transformation $dV'_{\rm unusual}=\gamma(1\pm\beta)dV$ which is different from the usual volume transformation $dV'_{\rm usual}=dV/\gamma$. The relation between both transformations reads $dV'_{\rm unusual}= dV'_{\rm usual}/(1\mp\beta).$ The quantities $(1\mp\beta)$ may be seen as correction factors for $dV'$ to be consistent with the equivalence of inertial frames in special relativity.

\section {Conclusion}
The absence of a convincing relativistic thermodynamics is a pending subject of special relativity. We can say that the relativistic thermodynamics is the missing chapter of relativity textbooks. In this state of affairs, we are free to explore alternative starting points to find a consistent formulation of relativistic thermodynamics. Here we have stressed the idea that meaningful relativistic temperature transformations in special relativity must satisfy the closure property that two successive temperature transformations are equivalent to a single temperature transformation with the corresponding reference frames velocities satisfying the parallel-velocity addition law of special relativity and  argued that this result is required by the equivalence of inertial frames in especial relativity. We have noted that the closure properties of both Planck-Einstein and Ott-Arzelies transformations of temperature involve a
a velocity addition law different from that of the special relativity and therefore these transformations are inconsistent with the equivalence of the inertial frames of special relativity. We have then suggested the relativistic temperature transformations $T'=\gamma (1\pm \beta) T$, which state that for $+\beta$ a body appears hotter (blueshifted temperature) and for  $-\beta$ appears colder (redshifted temperature). We have demonstrated that the transformations of the redshifted and blueshifted temperatures satisfy the same closure property of the Lorentz transformations in their standard configuration indicating that these temperature transformations are consistent with the equivalence of the inertial frames. By assuming the invariance of the general law of ideal gases and the Lorentz transformation for the volume, we have inferred the corresponding relativistic transformations for pressure. Analogously, by assuming the Lorentz invariance of entropy we have obtained the corresponding relativistic transformations for heat transfer and internal energy. We have claimed that the derived transformations  of temperature may be considered as particular cases of the so-called directional temperature.

\section*{Acknowledgement}
Thanks to our son Ricardo Heras for his interesting and instructive comments.

\section*{Data Availability Statement}
No data sets were generated or analysed in this paper.}


\newpage
\begin{thebibliography}{0}

\bibitem{1}
C.K. Yuen, Am. J. Phys. {\bf 38}, 246 (1970). \href{https://doi.org/10.1119/1.1976295}{https://doi.org/10.1119/1.1976295}

\bibitem{2}
D. Ter Haar and H.Wergeland, Phys. Rep. {\bf 1}, 31 (1971). \href{https://doi.org/10.1016/0370-1573(71)90006-8}{https://doi.org/10.1016/0370-1573(71)90006-8}

\bibitem{3}
P.T. Landsberg, Eur. J. Phys. {\bf 2}, 203 (1981). \href{https://doi.org/10.1088/0143-0807/2/4/003}{https://doi.org/10.1088/0143-0807/2/4/003}

\bibitem{4}
C. Liu, Stud. Hist. Phil. Sci. {\bf 25}, 983 (1994). \href{https://doi.org/10.1016/0039-3681(94)90073-6}{https://doi.org/10.1016/0039-3681(94)90073-6}

\bibitem{5}
I. Avramov, Russ, J. Phys. Chem, \textbf{77}, 179 (2003).

\bibitem{6}
D. Mi and H.Y. Zhong, Mod. Phys. Lett. A {\bf 24}, 73 (2009). \href{https://doi.org/10.1142/S0217732309026590}{https://doi.org/10.1142/S0217732309026590}

\bibitem{7}
T.K. Nakamura, Prog. Theor. Phys. {\bf 128}, 463 (2012). \href{https://doi.org/10.1143/PTP.128.463}{https://doi.org/10.1143/PTP.128.463}

\bibitem{8}
C.-Y. Wang, Adv. Nat. Sci. {\bf 6}, 13 (2013). \href{https://doi.org/10.3968/j.ans.1715787020120602.2121}{https://doi.org/10.3968/j.ans.1715787020120602.2121}

\bibitem{9}
J.J. Mares, P. Hubık, and V. Spicka, Fortschr. Phys., 1700018 (2017).
\href{https://doi.org/10.1002/prop.201700018}{https://doi.org/10.1002/prop.201700018}

\bibitem{10}
C. Farıas, V.A. Pinto and P.S. Moya, Sci. Rep. {\bf 7}, 17526 (2018).
\href{https://doi.org/10.1038/s41598-017-17526-4}{https://doi.org/10.1038/s41598-017-17526-4}

\bibitem{11}
Max Planck, In Sitzungsberichte der Königlich-Preussischen Akademie der Wissenschaften (Leipzig 1907) pp 542–570.

\bibitem{12}
M. Planck, Ann. Phys. Leipz. \textbf{26}, 1 (1908).
\href{https://onlinelibrary.wiley.com/doi/epdf/10.1002/andp.19083310602}{https://onlinelibrary.wiley.com/doi/epdf/10.1002/andp.19083310602}

\bibitem{13}
A. Einstein, Jahrb. Rad. Elec.\textbf{4}, 411 (1907).

\bibitem{14}
R. Penney, Nuovo Cim. A \textbf{43}, 911 (1966).
\href{https://doi.org/10.1007/BF02756369}{https://doi.org/10.1007/BF02756369}

\bibitem{15}
L. de Broglie, Diverses questions de m\'ecanique et de thermodynamique classiques
et relativistes, Lecture Notes in Physics Monographs book series, vol 32, Springer, Berlin, Heidelberg (1907). p146

\bibitem{16}
J.H. Fremlin, Nature \textbf{213}, 277 (1967). \href{https://doi.org/10.1038/213277a0}{https://doi.org/10.1038/213277a0}

\bibitem{17}
P.D. Noerdlinger, Nature \textbf{213}, 1117 (1967). \href{https://doi.org/10.1038/2131117a0}{https://doi.org/10.1038/2131117a0}

\bibitem{18}
J.H. Eberly, Nuovo Cim. B \textbf{48}, 167 (1967). \href{https://doi.org/10.1007/BF02712451}{https://doi.org/10.1007/BF02712451}

\bibitem{19}
W. Israel, Ann. Phys. \textbf{100}, 310 (1976). \href{https://doi.org/10.1016/0003-4916(76)90064-6}{https://doi.org/10.1016/0003-4916(76)90064-6}

\bibitem{20}
N. Agmom, Found. Phys. \textbf{7}, 331 (1976). \href{https://doi.org/10.1007/BF00711486}{https://doi.org/10.1007/BF00711486}

\bibitem{21}
G. Ares de Parga, B. L\'opez-Carrera and F. Angulo-Brown,  J. Phys. A \textbf{38}, 2821 (2005). \href{https://doi.org/10.1088/0305-4470/38/13/001}{https://doi.org/10.1088/0305-4470/38/13/001}

\bibitem{22}
Z.C. Wu, EuroPhys. Lett. \textbf{88}, 20005
(2009). \href{https://doi.org/10.1209/0295-5075/88/20005}{https://doi.org/10.1209/0295-5075/88/20005}

\bibitem{23}
A.S. Parvan, Ann. Phys. \textbf{401}, 130 (2019). \href{https://doi.org/10.1016/j.aop.2019.01.003}{https://doi.org/10.1016/j.aop.2019.01.003}

\bibitem{24}
H. Ott, Z. Phys. \textbf{175}, 70 (1963). \href{https://doi.org/10.1007/BF01375397}{https://doi.org/10.1007/BF01375397}

\bibitem{25}
H. Arzelies, Nuovo Cim. \textbf{35}, 792 (1965). \href{https://doi.org/10.1007/BF02739342}{https://doi.org/10.1007/BF02739342}
\href{https://doi.org/}{https://doi.org/}

\bibitem{26}
W.G. Sutcliffe, Nuovo Cim. \textbf{39}, 683 (1965). \href{https://doi.org/10.1007/BF02735833}{https://doi.org/10.1007/BF02735833}

\bibitem{27}
A. Børs, Proc. Phys. Soc. London \textbf{86}, 1141 (1965). \href{https://doi.org/10.1088/0370-1328/86/5/124}{https://doi.org/10.1088/0370-1328/86/5/124}

\bibitem{28}
A. Gamba, Nuovo Cim. B \textbf{41}, 79 (1966). \href{https://doi.org/10.1007/BF02783385}{https://doi.org/10.1007/BF02783385}

\bibitem{29}
T.W.B. Kibble, Nuovo Cim. B \textbf{41}, 72 (1966). \href{https://doi.org/10.1007/BF02711119}{https://doi.org/10.1007/BF02711119}

 \bibitem{30}
 C. Møller, Mat. Fys. Medd. Dan. Vid. Selsk. \textbf{36}, no. 1 (1967).

\bibitem{31}
L.A. Schmid, Nuovo Cim. B \textbf{47}, 1 (1967). \href{https://doi.org/10.1007/BF02712304}{https://doi.org/10.1007/BF02712304}

\bibitem{32}
M. Przanowski and J. Tosiek, Phys. Scr. \textbf{84}, 055008 (2011).
\href{https://doi.org/10.1088/0031-8949/84/05/055008}{https://doi.org/10.1088/0031-8949/84/05/055008}

\bibitem{33}
P.T. Landsberg, Proc. Phys. Soc. London \textbf{89}, 1007 (1966). \href{https://doi.org/10.1088/0370-1328/89/4/324}{https://doi.org/10.1088/0370-1328/89/4/324}

\bibitem{34}
P.T. Landsberg and K.A. Johns, Nuovo Cim. B \textbf{52}, 28 (1967).
\href{https://doi.org/10.1007/BF02710651}{https://doi.org/10.1007/BF02710651}

\bibitem{35}
P.T. Landsberg, Nature \textbf{214}, 903 (1967).
\href{https://doi.org/10.1038/214903a0}{https://doi.org/10.1038/214903a0}

\bibitem{36}
P.T. Landsberg and K.A. Johns, Proc. Phys. Soc. London A \textbf{306}, 477 (1968).
\href{https://doi.org/10.1098/rspa.1968.0162}{https://doi.org/10.1098/rspa.1968.0162}

\bibitem{37}
J. Lindhard, Physica \textbf{38}, 635 (1968).
\href{https://doi.org/10.1016/0031-8914(68)90011-6}{https://doi.org/10.1016/0031-8914(68)90011-6}

\bibitem{38}
N.G. van Kampen, Phys. Rev. \textbf{173}, 295 (1968).
\href{https://doi.org/10.1103/PhysRev.173.295}{https://doi.org/10.1103/PhysRev.173.295}

\bibitem{39}
G. Cavalleri and G. Salgarelli, Nuovo Cim. A \textbf{62}, 722 (1969).
\href{https://doi.org/10.1007/BF02819595}{https://doi.org/10.1007/BF02819595}

\bibitem{40}
H. Callen and G. Horwitz,  Am. J. Phys. \textbf{39}, 938 (1971).
\href{https://doi.org/10.1119/1.1986330}{https://doi.org/10.1119/1.1986330}
\bibitem{41}
D. Cubero et al., Phys. Rev. Lett. \textbf{99}, 170601 (2007).
\href{https://doi.org/10.1103/PhysRevLett.99.170601}{https://doi.org/10.1103/PhysRevLett.99.170601}

\bibitem{42}
N. Poplawski, (2021) arxiv:1902.05536
\href{https://arxiv.org/pdf/1902.05536.pdf}{https://arxiv.org/pdf/1902.05536.pdf}

\bibitem{43}
T.W. Marshall, Proc. Camb. Phil. Soc. \textbf{61}, 537 (1965).
\href{https://doi.org//10.1017/S0305004100004114}{https://doi.org//10.1017/S0305004100004114}

\bibitem{44}
F. Rohrlich, Nuovo Cim. B \textbf{45}, 76 (1966).
\href{https://doi.org/10.1007/BF02710587}{https://doi.org/10.1007/BF02710587}

\bibitem{45}
J.H. Eberly and A. Kujawski, Phys. Rev. \textbf{155}, 10 (1967).
\href{https://doi.org/10.1103/PhysRev.155.10}{https://doi.org/10.1103/PhysRev.155.10}

\bibitem{46}
 I.P. Williams, Nature \textbf{214}, 1105 (1967).
\href{https://doi.org/10.1038/2131118a0}{https://doi.org/10.1038/2131118a0}

\bibitem{47}
R. Balescu, Physica \textbf{40}, 309 (1968).
\href{https://doi.org/10.1016/0031-8914(68)90132-8}{https://doi.org/10.1016/0031-8914(68)90132-8}

\bibitem{48}
L.G. Taff, Phys. Lett. A \textbf{27}, 605 (1968).
\href{https://doi.org/10.1016/0375-9601(68)90074-1}{https://doi.org/10.1016/0375-9601(68)90074-1}

\bibitem{49}
R. Hakim and A. Mangeney, Lett. Nuovo Cim. \textbf{1}, 429 (1969).
\href{https://doi.org/10.1007/BF02752581}{https://doi.org/10.1007/BF02752581}

\bibitem{50}
I. Paiva-Veretennicoff, Physica \textbf{75}, 194 (1974). \href{https://doi.org/10.1016/0031-8914(74)90301-2}{https://doi.org/10.1016/0031-8914(74)90301-2}

\bibitem{51}
D. Eimerl, Ann. Phys. \textbf{91}, 481 (1975).
\href{https://doi.org/10.1016/0003-4916(75)90232-8}{https://doi.org/10.1016/0003-4916(75)90232-8}

\bibitem{52}
R.G. Newburgh, Nuovo Cim. B \textbf{52}, 19 (1979).
\href{https://doi.org/10.1007/BF02739036}{https://doi.org/10.1007/BF02739036}

\bibitem{53}
P.T. Landsberg, Phys. Rev. Lett. \textbf{45}, 149 (1980).
\href{https://doi.org/10.1103/PhysRevLett.45.149}{https://doi.org/10.1103/PhysRevLett.45.149}

\bibitem{54}
P.A. Goodinson and B.L. Luffman, Nuovo Cim. B \textbf{60}, 81 (1980).
\href{https://doi.org/10.1007/BF02723069 }{https://doi.org/10.1007/BF02723069 }

\bibitem{55}
W. Israel, Physica A \textbf{106}, 204 (1981).
\href{https://doi.org/10.1016/0378-4371(81)90220-X}{https://doi.org/10.1016/0378-4371(81)90220-X}

\bibitem{56}
R. Aldrovandi and J. Gariel, Phys. Lett. A \textbf{170}, 5 (1992).
\href{https://doi.org/10.1016/0375-9601(92)90382-v}{https://doi.org/10.1016/0375-9601(92)90382-v}

\bibitem{57}
P.T. Landsberg and G.E.A. Matsas, Phys. Lett. A \textbf{223}, 401 (1996).
\href{https://doi.org/10.1016/S0375-9601(96)00791-8}{https://doi.org/10.1016/S0375-9601(96)00791-8}

\bibitem{58}
P.T. Landsberg and G.E.A. Matsas, Physica A \textbf{340}, 92 (2004).
\href{https://doi.org/10.1016/j.physa.2004.03.081}{https://doi.org/10.1016/j.physa.2004.03.081}

\bibitem{59}
G.L. Sewell, J. Phys. A \textbf{41}, 382003 (2008).
\href{https://doi.org/10.1088/1751-8113/41/38/382003}{https://doi.org/10.1088/1751-8113/41/38/382003}

\bibitem{60}
G.L. Sewell, Rep. Math. Phys. \textbf{64}, 285 (2009).
\href{https://doi.org/10.1016/S0034-4877(09)90033-7}{https://doi.org/10.1016/S0034-4877(09)90033-7}

\bibitem{61}
G.L. Sewell, J. Phys. A \textbf{43}, 485001 (2010).
\href{https://doi.org/10.1088/1751-8113/43/48/485001}{https://doi.org/10.1088/1751-8113/43/48/485001}

\bibitem{62}
G.W. Ford and R.F. O’Connell, Phys. Rev. E \textbf{88}, 044101 (2013).
\href{https://doi.org/10.1103/PhysRevE.88.044101}{https://doi.org/10.1103/PhysRevE.88.044101}

\bibitem{63}
L. Gavassino, Foundations of Physics \textbf{50} 1554 (2020).
\href{https://doi.org/10.1007/s10701-020-00393-x}{https://doi.org/10.1007/s10701-020-00393-x}

\bibitem{64}
C.V. Heer and R.H. Kohl, Phys. Rev. \textbf{174}, 1611 (1968).
\href{https://doi.org/10.1103/PhysRev.174.1611}{https://doi.org/10.1103/PhysRev.174.1611}

\bibitem{65}
P.J.E. Peebles and D.T. Wilkinson, Phys. Rev. \textbf{174}, 2168 (1968).
\href{https://doi.org/10.1103/PhysRev.174.2168}{https://doi.org/10.1103/PhysRev.174.2168}

\bibitem{66}
R.N. Bracewell and E.K. Conklin, Nature \textbf{219}, 1343 (1968).
\href{https://doi.org/10.1038/2191343a0}{https://doi.org/10.1038/2191343a0}

\bibitem{67}
G.R. Henry et al., Phys. Rev. \textbf{176}, 1451 (1968).
\href{https://doi.org/10.1103/PhysRev.176.1451}{https://doi.org/10.1103/PhysRev.176.1451}

\bibitem{68}
D. van Dantzig, Physica \textbf{6}, 673 (1939).
\href{https://doi.org/10.1016/S0031-8914(39)90072-8}{https://doi.org/10.1016/S0031-8914(39)90072-8}

\bibitem{69}
 T.K. Nakamura, Phys. Lett. A \textbf{352}, 175, (2006).
\href{https://doi.org/10.1016/j.physleta.2005.11.070}{https://doi.org/10.1016/j.physleta.2005.11.070}

\bibitem{70}
T.K. Nakamura, EPL \textbf{88}, 20004 (2009). \href{https://doi.org/10.1209/0295-5075/88/20004}{https://doi.org/10.1209/0295-5075/88/20004}

\bibitem{71}
Z.C. Wu, EPL \textbf{88}, 20005 (2009).
\href{https://doi.org/10.1209/0295-5075/88/20005}{https://doi.org/10.1209/0295-5075/88/20005}

\bibitem{72}
J.M. L\'evyLeblond, Am. J. Phys. \textbf{44}, 271 (1976).
\href{https://doi.org/10.1119/1.10490}{https://doi.org/10.1119/1.10490}

\bibitem{73}
T.S. B\'{\i}r\'{o} and P. V\'{a}n, EPL, \textbf{89} 30001 (2010).
\href{https://doi.org/10.1209/0295-5075/89/30001}{https://doi.org/10.1209/0295-5075/89/30001}

\bibitem{74}
T.S. B\'{\i}r\'{o}, {\it Is There a Temperature?} (Springer, NY, 2011).

\bibitem{75}
J.S. Lee and G.B. Cleaver, Mod. Phys. Lett. A  \textbf{30}, 1550045 (2015). \href{https://doi.org/10.1142/S0217732315500455}{https://doi.org/10.1142/S0217732315500455}

 \bibitem{76}
L. Gavassino, Foundations of Physics \textbf{52}, 11 (2022). \href{https://doi.org/10.1007/s10701-021-00518-w}{https://doi.org/10.1007/s10701-021-00518-w}

 \bibitem{77}
J.J. Mares et al, Physica E \textbf{42}, 484 (2010). \href{https://doi.org/10.1016/j.physe.2009.06.038}{https://doi.org/10.1016/j.physe.2009.06.038}

 \bibitem{78}
W. M. Haddad, Entropy \textbf{19}, 621 (2017). \href{https://doi.org/10.3390/e19110621}{https://doi.org/10.3390/e19110621}

\end{thebibliography}
\end{document}